# Virtual Simulation Objects Concept as a Framework for System-Level Simulation

Sergey V. Kovalchuk, Pavel A. Smirnov
e-Science Research Institute,
National Research University of IT, Mechanics and Optics
St. Petersburg, Russia
kovalchuk@mail.ifmo.ru

Sergey S. Kosukhin, Alexander V. Boukhanovsky
e-Science Research Institute,
National Research University of IT, Mechanics and Optics
St. Petersburg, Russia

*Abstract* — **This paper presents Virtual Simulation Objects (VSO) concept which forms theoretical basis for building tools and framework that is developed for system-level simulations using existing software modules available within cyber-infrastructure. Presented concept is implemented by the software tool for building composite solutions using VSO-based GUI and running them using CLAVIRE simulation environment.**

## I. Introduction

Contemporary e-science tasks often require large amount of computer simulations to be performed. The simulation process today is often characterized not only by computational complexity, but also by structural complexity: solving the e-science tasks requires composition of resources from different classes – hardware, software, informational, even human resources. All these resources should be integrated and managed in the appropriate way to solve particular complex simulation tasks. The issue of integration resources within one composite solution becomes more complex as there is a great diversity of resources to be integrated.

There exists a lot of simulation software already developed in almost every problem domain. Some of them are rather modern, but some are developed decades ago and still are in use as they have earned the trust of domain specialists. Considering integration issues both legacy and modern software pieces are to be integrated within composite solutions. One of the common ways for collecting the software within one environment is offered by problem solving environment (PSE) concept [1]. But still there are two problems which occur while solving the integration issue. Firstly, taking into account huge diversity in technologies, execution platforms, and data formats etc. it is still a great challenge to build large composite application even within a PSE. Secondary, there are difficulties related to the use of the existing third-party software with a lack of knowledge about its functionality, and internal features.

Looking at contemporary computational resources we can also see high level of diversity in architectures, technologies, supported software etc. Moreover today we have an ability to combine different computational resources using such approaches as metacomputing, Grid [2], or Cloud [3]. In this case the problem of integration becomes more important as we should take care of performance issues in heterogeneous computational environment because of computational intensity of e-Science software.

Today common approach for building composite solutions using diverse resources is based on the workflow structure [4] which allows calling integrated software pieces within one algorithm using software-as-a-service (SaaS) approach. But contemporary trends in e-science show interest to more system-based approach for building investigation process (see e.g. [5]): instead of running several procedures arranged within workflows, we probably need to make a shift to exploring the whole system within the simulation-driven approach. This issue becomes much more topical when interdisciplinary tasks are concerned. This kind of tasks requires new and more complex approaches, which need knowledge from many problem domains to be combined within one solution.

Thus the mentioned paradigm shift leads us to more domain-specific and more general way of system analysis than the workflow-based approach. This may be claimed because the investigated system and its properties belong to the problem domain, while workflow in fact is the object much more related to IT-domain. Moreover as the end-users of e-science solutions are domain specialist whose interest is exploration of domain-specific system, the new approach may lead us to more user-friendly solutions and platforms. Nevertheless taking into account the mentioned issues of resource integration and composite solution building, the new approach to building composite solution is required. This approach should a) allow domain-specialists to build the interactive model of system being explored; b) run the simulation for getting required parameters as a results.

In this paper we present the Virtual Simulation Object (VSO) concept which is devoted to manage the mentioned issues, by using domain expert knowledge for a) system-level model construction and b) running simulation process in a hidden way using the defined model. This concept is proposed as an extension to Intelligent PSE (iPSE) concept [6], which offers the domain-specific knowledge-based approach to use cyber-infrastructure for running simulation. iPSE concept and platform based on it are developed and used by e-Science Research Institute during several projects within last years. The remaining structure of the paper is as follows. In section II the

Virtual Simulation Objects concept is described with more details. Section III shows implementation of the concept as an extension to CLAVIRE platform, which is based on iPSE concept. Section IV presents discussion on features of VSO and related fields of knowledge. Section V claims conclusions on VSO concept and current state of its implementation.

## II. VIRTUAL SIMULATION OBJECTS

### A. Conceptual Requirements

The main idea behind VSO concept is to develop an approach for domain-specific system's description, which allows to run simulation and to use their results as characteristics of the explored system. Within VSO concept system is described as a set of objects, which interact with each other. Each of the VSOs is related to some real-world object, which forms a real system. So the set of VSOs can be considered as an image of system to be investigated using simulation. This approach should response to the following requirements:

1) *System's description* should be considered as a structural model of investigated system, containing the interactive objects. Each object is described by a set of characteristics, which are simulated by a set of interconnected models. Each model can be implemented by a composite application, which include calling of particular software. System description itself can be processed as VSO, so the VSO entity can be considered as a hierarchical structure. System description should allow to manage the simulation process, performed using resources within the available cyber-infrastructure.

2) VSO should contain knowledge to support interconnection with other VSOs and to run simulation within composite solution automatically. So VSO needs to be a *distributable set of knowledge* which can be integrated within processing system to make it support simulation of particular objects. To do that the set of knowledge should include following subsets: a) set of domain-specific knowledge, which defines used simulation models, object input and output parameters; b) set of technological knowledge, which allows performing simulation in an automatic way; c) set of task-related knowledge, which allows to process the integrated composite solution, defined in relation to system's description. The goal of these knowledge sets' usage is to make system's description *interpretable* in two ways: a) the system should have sense within a problem domain, so the user having the knowledge within this domain should intuitively understand the structure and its usage; b) the description should be machine-interpretable so the simulation process can be performed in an automatic way.

3) Typical e-science task consists of three stages: *modeling, simulation and result analysis*. VSO concept should present continuous technological and informational support for all three stages of this process. System description should be core structure for performing all three stages: define system's structural mode by selections VSOs, tune their parameters and interconnect them; automatically perform simulation process according to the defined structure and provided data; present the simulated data available for analysis and its visualization arranged with the system's structure. The better way to do this is to give the system's structure a graphical interactive representation, which allows user to perform all this stages.

### B. Usage of Knowledge

VSO concept would be impossible without strong usage of expert knowledge. There are three main domain of knowledge involved into solving e-science problems: problem domain, IT domain and general problem solving domain. Within iPSE concept [6] the first two classes of knowledge are formalized. That allows providing automatic processing of tasks expressed using abstract workflows (AWF), which describes the workflows with specification of software packages, domain-specific methods and parameters. With the use of knowledge AWF is translated into concrete workflow (CWF), which defines calling particular services using input data compiled with the provided domain-specific parameters. This approach uses a) knowledge on parameters of available software and theirs domain meaning; b) knowledge about performance of available software running on different hardware architecture (that allows to tune running parameters to gain better performance); c) knowledge about available platforms and running mode (this part of knowledge allows to adopt the running parameters to the particular hardware); d) collection of the best practices of running software to provide the user with the prepared patterns for solving of different domain-specific tasks.

There exist the extensions for iPSE concept which are oriented to the use of the third knowledge domain – knowledge which allows solving domain-specific problems with general approach. We use the following conceptual hierarchy to describe simulation process:

*1) Simulated object*, which represents the main entity, which is explored during simulation. The object can be concerned as a composite entity, or system of objects. In this case the simulation process might be defined for whole composition or for separate objects, with explicit definition of theirs interaction. The explored system can be concerned as a composition of the objects.

*2) Simulated model*, which describes a set of static and dynamic characteristics of the object and can be used to explore it. The model can be defined as static if it describes object's structure within a fixed moment of time. Models of this type can be used for structural analysis of the system. In case the model describes evolution of the system within the time domain it is defined as dynamic.

*3) Method* can be defined as imperative description of the model usage process. With the given input parameter set, the method allows to calculate output model parameters. Methods are implemented in simulation software as algorithms for solving some particular domain problem. Considering this hierarchy the problem can be defined as a composition of object description and model.

*4) Software packages* are used as implementations of the defined methods in form of an algorithm. Usually this kind of software is developed by the domain specialists. Often there is a huge amount of software with different variants of the same method implementation.

*5) Service* within a distributed computational environment (in case we are using SaaS approach) can be considered as the software deployed on computational resource (hardware or virtual machine). Services are the low-level elements of regular workflows.

Most modern distributed simulation environments work on level 5 (services) of this hierarchy. More advanced environments and PSEs get to the level 4 (software packages). General iPSE concept covers level 3 (methods) which allows to call domain-specific methods within AWFs. One of the extensions to iPSE concept [7] allows to compare and to select models and methods according to the quantitative domain-specific quality estimation based on the analysis of provided data and interactive dialog with the user. So this extension allows operating on levels 2 and 3. Concerning this hierarchy VSO concept is developed as the conceptual extension which works on levels 1 to 3. Thus, the knowledge used within this extension should contains information on the explored objects, related models and should be linked to the lower levels of presented conceptual hierarchy to make a simulation running available to the user.

Within VSO concept we define the following structures, which are used to represent knowledge within ontology structure:

*1) Virtual Simulation Objects* themselves. This structurual entity is used to organize simulation models and to define structure of the explored system's virtual representation.

*2) VSO data*, which can be defined by the user, or obtained by running the simulations. It can be devided into subgroups according to the processing style: a) *bases* – domains for defining other parameters of object; b) *parameters* – data that define the objects (doesn't change during the simulation); c) *values* – data processed during the simulation.

*3) Simulation models*, related to defined VSO, which can use VSO data as the input and output values. Models as well as data should be tunable by switching on/off and defining some additional options.

*4) Simulation scenario*, define the variations of models usage, which can differ by the input and output sets as well as the options set. Thus, together with the implemented method, the simulation scenario define relations between the model and available values within VSO.

*5) Scenario implementation* links the simulation scenario with available software using domain-specific parameters. Parameters of software running can be either set directly within the scenario, or defined as a model-level option, or obtained as a value within VSO. This piece of knowledge define the parts of composite simulation software to be run using iPSE-based environment. Thus, the scenario implementation can be defined as a parts AWF.

The entities described above define the structure of VSO which will be described in the next section.

*C. Virtual Simulation Obects Structure*

The structure of virtual simulation object can be considered as a tuple:

$$VSO = \langle B, V, Q, M, E \rangle, \qquad (1)$$

where $B$ is a set of available bases; $V$ is a set of values, which can be defined on the bases from $B$; $Q$ is a set of quality metrics for values from $V$; $M$ is a set of models, which operates with values from $V$; $E$ is a set of interconnections between models. Elements of virtual simulation objects are described further with more details.

*Bases.* Bases can be defined as a parameters domain for values. Typical examples here are space, time and groups, which can make different combinations [8]. Considering VSO $B$ is defined as a set of available positions within the used bases combination. E.g. if we explore dynamic of the sea level we can define $B$ as a set of two-dimensional grids, used within the simulation process, considered at every time step (so we have a combination of time and space bases). Bases within $B$ can be defined absolutely or relatively. E.g. in the previous example space grids can be defined using absolute coordinates, while time can be counted from any $t_0$. Bases describe the object configuration in a form which is used during simulation process.

*Values.* Values are any quantified entity (scalar or vector), which either can be associated with the elements from $B$ set or can be an abstract values defining object's characteristics. E.g. the sea level can be defined on a particular grid within defined moment of time. Values set can be decomposed into two sets: $V = V_{CONST} \cup V_{VAR}$, where $V_{CONST}$ are invariant parameters, which describe the virtual object itself, $V_{VAR}$ is a set of variable parameters, which is processed within the simulation process.

*Quality metrics.* This set defines the available meta-characteristics of values. E.g. it can be defined as $Q_M \times Q_{Expert}$, where $Q_M = \{0,1\}$ shows if the value is measured or simulated, $Q_{Expert} = \Re$ – is the quantitative quality estimation defined by the experts. The quality metrics could be related to the quality estimations of the solutions proposed within extended iPSE concept [7].

Using defined elements a set of possible data structures can be defined as follows:

$$D = V \times (B \cup \varnothing) \times Q \qquad (2)$$

*Models.* A set of models $M$ can be defined as follows:

$$M = \{m = \langle D_{IN}, D_{OUT}, P, S, sp, sv \rangle\} \qquad (3)$$

where $D_{IN}, D_{OUT} \subset D$ define required input and available output for the model; $P$ is a set of available software packages which implements defined models; $S$ is a set of available scenarios containing calls of the software packages, which can be defined using a function $sp: S \to P^*$; $sv$ is a function which defines extra parameters required by the scenario $sv: S \to D^*$. Extra parameters can be required either by scenario itself or by packages used within the scenario.

*Interconnection.* Models can be interconnected using the transition edges $e \in E$, which can be defined as follows:

$$e = \langle m_1, m_2, d \rangle : d \in m_1.D_{OUT} \cap m_2.D_{IN}, \quad (4)$$

where $m_1$ and $m_2$ are the interconnected models; $d$ is a data structure which is used to interconnect the models. Thus $\langle M, E \rangle$ defines a graph structure which can be used to perform a simulation process during the interpretation.

We define the following composition operator which allows constructing the composite VSOs:

$$VSO_C = VSO_1 \bullet VSO_2 \quad (5)$$

where $VSO_C$ is composed according to the following rules:

$$\begin{aligned} B_C &= B_1 \cup B_2, \\ V_C &= V_1 \cup V_2, \\ Q_C &= qm(Q_1, Q_2): Q_1 \leftrightarrow Q_C, Q_2 \leftrightarrow Q_C, \\ M_C &= M_1 \cup M_2 \cup M_T, \\ E_C &= E_1 \cup E_2 \cup E_T. \end{aligned} \quad (6)$$

Here $qm$ is a function which merge quality metrics in a way that unified metrics set $Q_C$ can be mapped on sets $Q_1$ and $Q_2$ and vice versa. $M_T$ is a set of transition models, which allows to transfer the similar values from one object to another:

$$M_T = \begin{cases} \langle D_{IN}^{(T)}, D_{OUT}^{(T)}, P_T, S_T, sp_T, sv_T \rangle : \\ D_{IN}^{(T)} = \{\langle v, b_1, q_1 \rangle \in D_1\}, \\ D_{OUT}^{(T)} = \{\langle v, b_2, q_2 \rangle \in D_2\}, \\ v \in V_1 \cap V_2, sv_T : S_T \to V_1 \cup V_2 \end{cases}. \quad (7)$$

In some cases it is possible to apply transition models without calling any external packages (i.e. $P_T = \varnothing$). E.g. simple selection of the value can be considered as a transition model. Transition models set allows to define transition edges set:

$$E_T = \begin{cases} \langle m, m_T, d \rangle : \forall d \in m_T.D_{IN} \cap m.D_{OUT} \} \cup \\ \langle m_T, m, d \rangle : \forall d \in m_T.D_{OUT} \cap m.D_{IN} \} \end{cases}. \quad (8)$$

Interpretation of VSO structure consists of three main stages:

*1) Selection of VSO structure.* Graph structure $\langle M, E \rangle$ can be filtered according to a particular task which is to be solved. Some of models and edges are excluded from structure either by the user or by intelligent support system.

*2) Data processing.* Selected graph is processed using a given data set to get the requested data set (both of them are a subsets of $D$). During this processing the system should find and analyze all the possible sub-graphs which define the models' usage process. The final choice can be performed either by the user or by intelligent support system.

*3) Simulation processing.* Selected sub-graph can be used to construct a composite application describing data processing and running of software packages.

### D. Visual Representation of Virtual Simulation Objects

As it was mentioned before, the interactive graphical representation is one of the requirements for developed VSO concept, because the concept should form a basis for interactive simulation tools. Thus, the basic schema for VSO visual representation was developed. Fig. 1 shows an example of two interacting VSO: the sea and the ship placed on this sea.

Each visual representation of VSO consists of the following parts:

*1) Header.* Represents the naming of the object and allows to tune the metaparameters of VSO processing (inluding mode, which is defined by the scenario of general task solving: analysis, forecast, optimization etc.)

*2) Bases pane.* Allows tuning of bases parameters (e.g. define spatial grids, forecast period etc.)

*3) Object parameters pane.* Defines a set of parameters which define the explored object.

*4) Models graph.* Shows the structure of interconected models and values within the object.

Each dataset block (in object parameter pane or in models graph) is linked with some of available basis defined in the bases pane. User can select if the data is generated by simulation (received from some model), downloaded from any available data storage or defined by the user. Data blocks can be marked with one of the following states: "OK" – the dataset is checked for availability and correctness (see "Near-water wind" input dataset on the fig. 1); "?" – the dataset need to be defined and/or tuned (see "Bathymetry" dataset, which isn't presented at the moment); "X" – the dataset which will be unavailable during the simulation process (see "Wave parameters" dataset, which will not be produced due to "Spectrum parameterization" model is disabled). Options for tuning data integration process are available by "…" button. Particular set of options is defined by the role of dataset within the VSO.

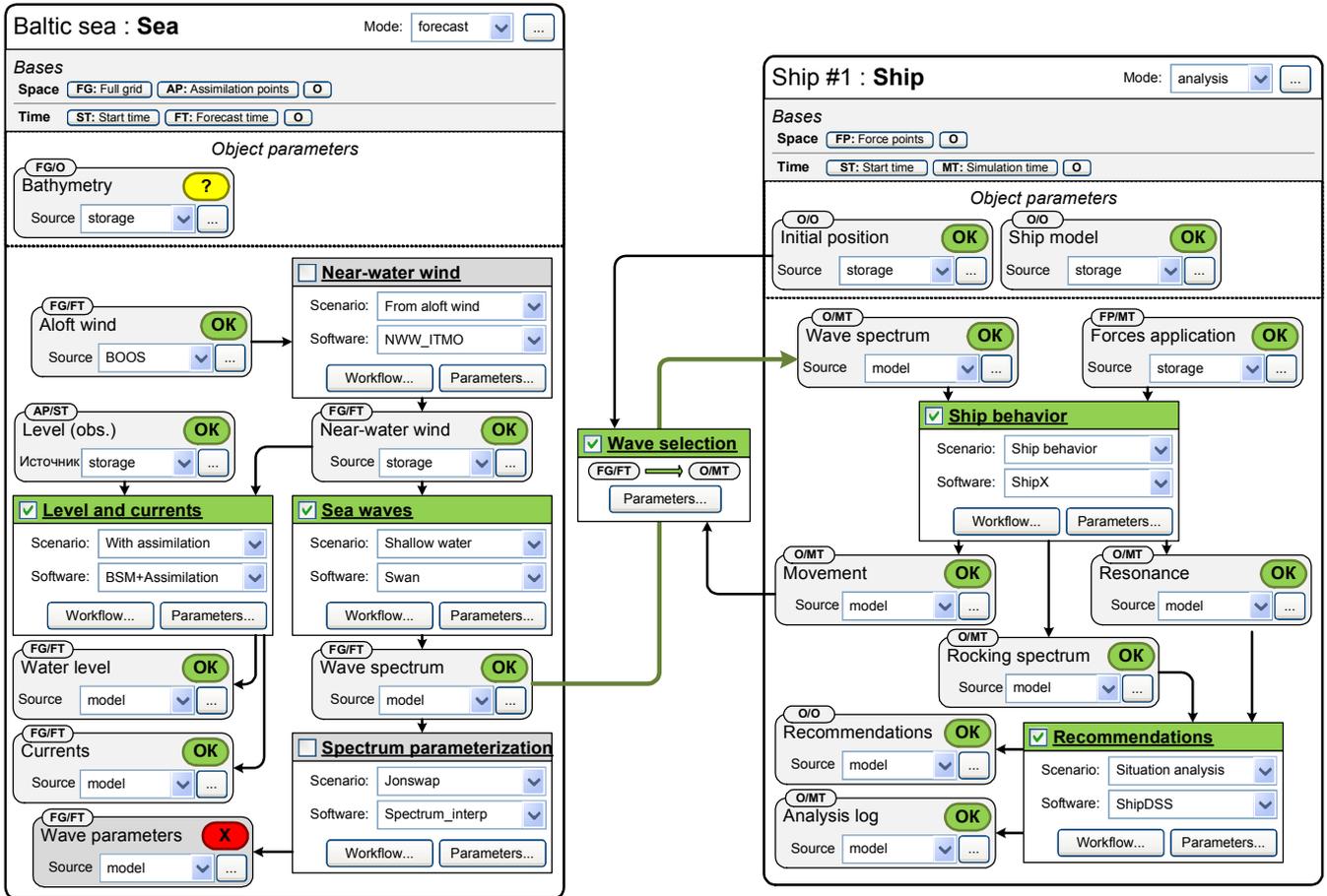

Figure 1. Virtual simulation objects graphical representation

Each model block contains the related sets of scenarios and their implementations using the available software packages. A set of scenarios define the possible relations between models. Besides the selection of scenario and its implementation the user can tune additional options, which may change according to selected scenario and implementation. Also the user should have ability to select the models of interest according to his/her task. This selection is especially important in case of interdisciplinary tasks – it is possible that only a few models are required to simulate object's behavior with the required quality. So, on the fig. 1 models "Level and currents", "Sea waves", "Ship behavior" and "Recommendation" are selected within two objects. Using "Near-water wind" and "Level (obs.)" datasets "Recommendations" is produced as a result of simulation.

Interconnection between objects is done by presenting the transition models, which allow to map the values between bases of two or more objects. These models can use existing software packages or can be presented by the scripts within VSO environment. Transition models are also tunable with additional options set as well as regular models. On the fig. 1 data ("Wave spectrum") is transferred from one object to another through the selection procedure (model) which allows to map space and time bases of one object (full grid and forecast time) on the corresponding bases of another object (current location and simulation time).

*E. Virtual Simulation Object Management System*

VSO management system is a software environment which supports a) user interaction according to VSO visual representation approach; b) VSO-based work within simulation process; c) presentation of simulation and observation results and support of the user with arrangement to the VSO structure. Thus, VSO management system should tightly interact with several software systems (see fig. 2):

*1) iPSE-based simulation environment.* As VSO is an extension of iPSE concept, its implementation should be developed in tight connection with iPSE-based simulation environment. Interaction with this environment is performed within two main direction: a) simulation tasks constructed during interpretation of VSO structure are tranmitted to the simulation enviroment in form of AWFs; b) data, processed by iPSE-based environment (collected as the simulation results or received from the external sources) are transferred to VSO management system to be displayed with the arrangement according to the to system's object structure. Observation of the real system can be consumed by the simulation environmend (and thus, by the VSO management system) among other data sources.

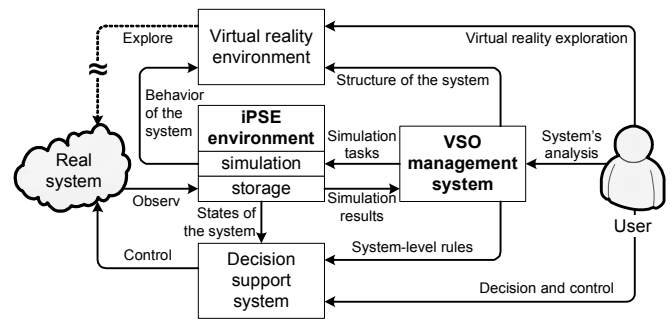

Figure 2. VSO management system

*2) Virtual reality environment.* VSO system's structure corresponds to some real-world sytem, which can be represented within the virtual environment, presented to the user. Bases datasets (spatial, temporal, group) can be used for the correct placement of virtual objects. Simulation results available within iPSE environment can be used to show behavior of virtual system. Thus, interactive virtual environmen, connected to VSO management system and iPSE-based simulation enviroment, allows to construct the virtual scenes in an automatic way. Also the virtual reality environment may use special hardware (3d walls, caves etc.) to be more expressive to the user. The virtual reality system in this case can be considered as a powerful tool for the simulation-based exploration of the real-world systems.

*3) Decision Support System.* Interaction between DSS and VSO management system allows the additional arrangment within the decision and control processes. Rule set of DSS as well as the results of simulation (which can be used as the decision arguments) can be linked with the parts of system's description within VSO framework and thus domain-specific decision and control solutions can be build in the unified way.

*F. System-Level Issues*

As VSO concept is built around the domain-specific system investigation, the system's description plays an important role within the concept. In this section some issues related to system's description are mentioned.

**Knowledge distribution using VSO.** As it was mentioned VSO can be considered as a unit of knowledge distribution. Experts from different problem domains can share their knowledge on simulation properties of any object using available software in the separate classes of VSO. This is a thing of especial interest in case of interdisciplinary researches being performed.

In case of sharing knowledge several issues should be taken into account:

*1) Automatic integration.* If several objects within the system's structure are developed by different researches, VSO management system should allow easy-to-use creation of transition models. This becomes possible by developing a general approach of basis processing due to the fact that there are not so much general classes of bases (i.e. space, time and group).

*2) Unification of knowledge.* To make VSO interconnected automatically it is practical to use the existing high-level ontologies (like SUMO, CYC, YAGO etc.) as they are trying to develop the description of whole universe as a set of related entities. E.g. concerning the example on Fig. 1 in relation to SUMO [9] ontology it is possible to link Ship and Sea objects to WaterVehicle (from Transportation ontology extension) and WaterArea (from Geography ontology extension) ontology terms respectively. Thus, the relations between the terms within ontology (using Transportation process) can produce relationship between VSOs.

*3) VSO editor.* To make the things easier to use, GUI-based editor of VSO classes should be developed. Developed classes then may be distributed and integrated into VSO management system. Instances of these classes are used to construct the system's description.

**System investigation tasks.** There are different classes of tasks which can be solved using VSO approach. These classes can be used to arrange simulation of objects within the system's structure according to high-level template. E.g. classes can be as follows: a) analysis – given the set of properties of the system, it is needed to estimate several other properties; b) forecast – given the initial state of the system, it is required to estimate state of the system in a future; c) optimization – given the boundary condition and quality function, it is required to find optimal input parameters according to the function. All these classes can produce special simulation procedures which finally form AWF to be run and response to be returned to the user. Thus the class of the task should be defined before VSO-based simulation run (see field Mode on Fig. 1).

III. IMPLEMENTATION DETAILS

To put the presented concept into practice, the implementation of VSO management system software is now being developed. It is integrated with CLAVIRE (CLoud Applications VIRtual Environment) platform which allows building composite the applications with the use of the set of the domain-specific software available within the service-oriented distributed computational environment. CLAVIRE environment is based on iPSE concept. It was developed during several projects, performed by e-Science Research Institute over last few years.

*A. Solution's Structure*

Structure of developing VSO management system is presented on Fig. 3. Most software modules are implemented using Microsoft .NET Framework. Main modules of this implementation are as follows:

*1) VSO core service* is the main module which implements

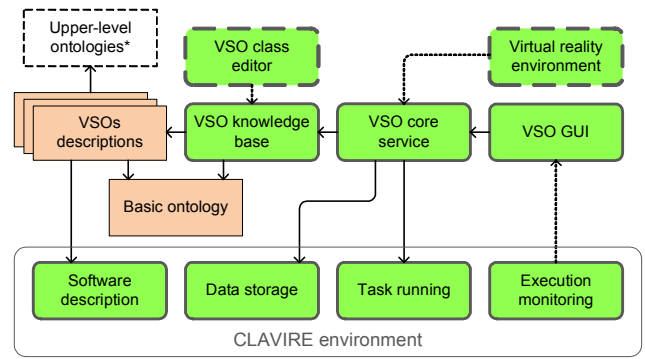

Figure 3. VSO management system implementation

the logic of presented concept. It is implemented as WCF-service which provides other modules with the access to the knowledge database and performs main processes using VSO structure (runs simulation, manage data etc.). It calls the corresponding services within CLAVIRE environment which in turn are also implemented using service-oriented approach.

*2) VSO knowledge base* is a library which manage the access to the structures that represent VSO classes. Set of VSO classes is presented as the ontology structure using OWL language. This ontology includes several descriptions of VSO classes (each in separate owl-file) linked to a) basic ontology, which defines VSO concepts; b) upper level ontology, which is used to linking VSOs. Knowledge base is linked to the software description presented within CLAVIRE environment as PackageBase service.

*3) VSO class editor* is a web-application which allows to add and modify VSO classes description stored in the knowledge base.

*4) VSO GUI* is a web-application (developed using the Silverlight technology) which implements the visual representation concept described earlier. The implemented visual representation can be linked to monitoring service of CLAVIRE environment to provide the user with information on simulation running process.

*5) Virtual reality environment* is a software solution for representing simulation results within the virtual space. It can use special hard- and software: e.g. during several experimentations the virtual reality software was run using 3D-wall hardware allowing to use stereoscopic 3D effect.

*B. Usage Example*

To test the implemented system the ship behavior during sailing in the sea was simulated. Fig. 4 shows different software modules involved into the simulation process. Fig. 4a represents the main user interface of VSO GUI web-application during process of VSO tuning: Sea and Ship objects are selected and tuned. Sea object is responsible for simulation of waves using SWAN software. Characteristics of simulated waves' field are used then to simulate behavior of the ship: parameters of rocking and expert system estimation of danger level for the ship. Simulation is run using CLAVIRE environment which is provided with AWF with blocks describing corresponding software running. CLAVIRE user interface (which is web-application as well) with this AWF opened is presented at Fig. 4b. Screenshots on Fig. 4c shows virtual reality environment that is running ShipXDS software (developed by e-Science Research Institute). This software can display the movement of the ship over the sea surface along with the additional debug information. This software can be run using simple PC or special hardware supporting stereoscopic 3D effect.

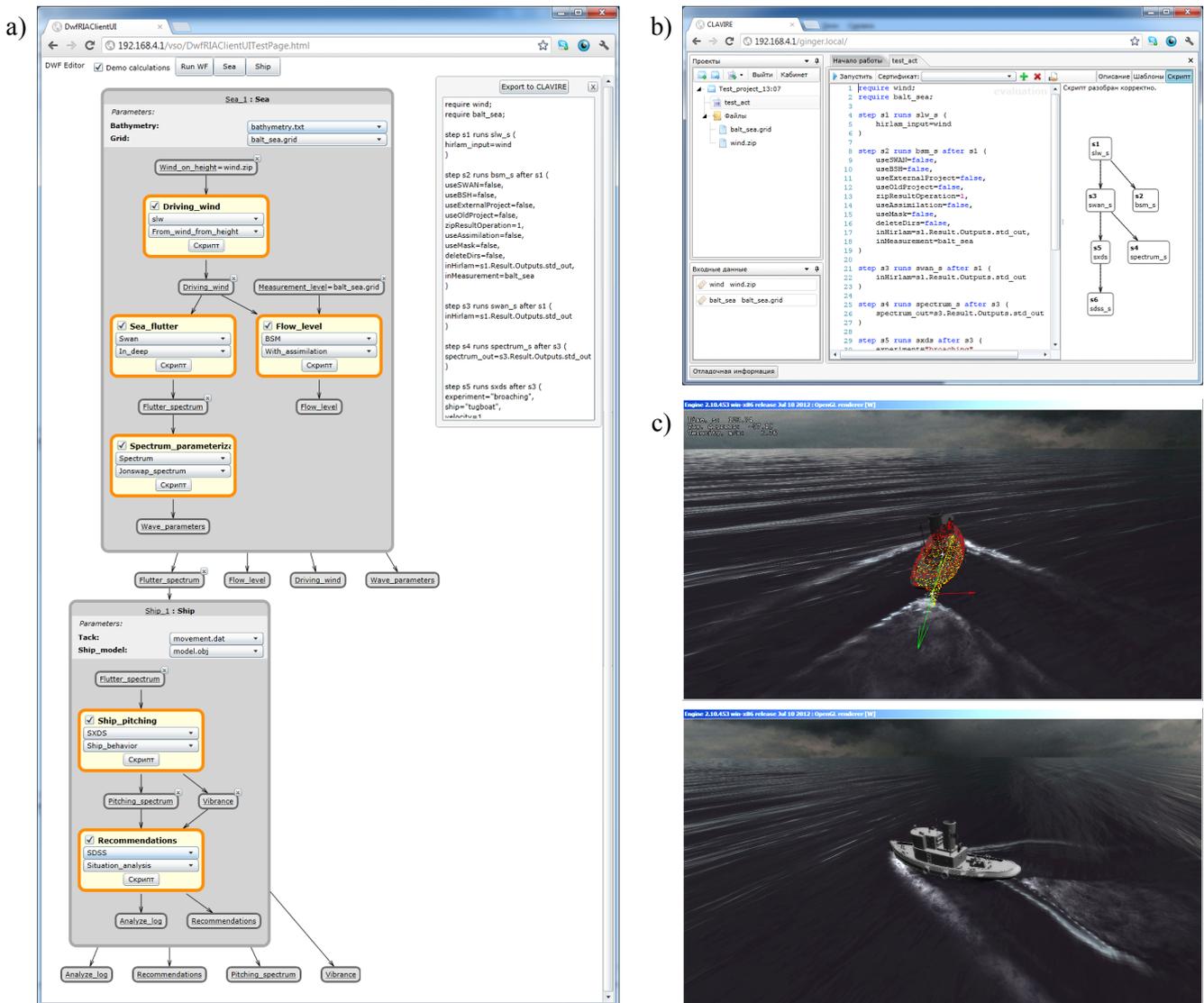

Figure 4. Running simulation using VSO management system

The implementation of VSO management system is now under development. Thus, it shows only partial functionality of full VSO management system described earlier. Nevertheless it demonstrates the potential capabilities of VSO approach within simplification trends and shifts toward domain-specific human-computer interaction.

## IV. DISCUSSION

Presented VSO concept is called to try fill in the gap between three large fields of knowledge: information technologies – which is used to have access to large amount of hardware and software resources within one composite application; knowledge engineering – which gives ability to manage diverse collections of knowledge bases; and theory of modeling and simulation along with general systems theory – which define general approaches to manage virtual system's description during solving simulation tasks. This conjunction of three different fields allows building the powerful automatic support tools for solving complex e-science tasks related to computer simulation. Today there are a lot of works related to these fields separately or in couples. E.g. there are several projects trying to bring the knowledge-based and intelligent support into the distributed simulation software using ontologies, workflow patterns, AI algorithms (see [10-12] for example). On the other hand there are huge amount of works started a long time ago devoted to development of simulation theory along with theoretical and software frameworks for solving system level tasks using simulation (e.g. see great works of Klir [8], Zeigler [13] and other [14, 15]). So today when each of these fields is strong enough and there exists a certain interest towards the system-level view within the simulation-based approach [5] it is time to join these fields.

On the other hand, another triplet may be introduced to support the presented idea. VSO concept is developed for the continuous support of modeling, simulation and data analysis. Nowadays there are frameworks and standards for solving these tasks separately (see for instance SysML [16] for modeling and HLA [17] for simulation). Presented approach is developed to join all mentioned processes around the structural model of investigated system. Also that corresponds to model-based systems engineering which is considered within development trends in a field of systems engineering [18].

## V. CONCLUSION

This paper presents the Virtual Simulation Objects (VSO) concept as a theoretical and applied framework to support simulation-based investigation of complex systems on domain-specific level. Proposed approach allows managing simulation process using the system's structural model, which can be presented to the user via interactive graphical representation. That structural model allows using redistributable knowledge sets presented as description of virtual simulation objects that represents the elements of real-world system which can be simulated using available software. The concept is now being implemented as a graphical web-application extending the existing CLAVIRE simulation environment.

ACKNOWLEDGMENT

This work was supported by the project granted from Decree 218 of Government of the Russian Federation under contract 13.G25.31.0029 and project granted from the Leading Scientist Program (Decree 220) of the Government of the Russian Federation under contract 11.G34.31.0019.